\documentstyle[10pt,epsfig]{article}
\newcommand{\ts}{\tilde{s}}

\newcommand{\beq}{\begin{equation}}
\newcommand{\eeq}{\end{equation}}
\newcommand{\beeq}{\begin{eqnarray}}
\newcommand{\eeeq}{\end{eqnarray}}

\def\({\left(}
\def\){\right)}

\def\mpis{m^2_\pi}

\def\wf{wave function }
\def\wfs{wave functions }
\def\ff{form factor }
\def\ffs{form factors }

\begin{document}
\renewcommand{\thefootnote}{\fnsymbol{footnote}}

\begin{flushright}
{\Large hep-ph/9503473}
\end{flushright}

\begin{center}
{\LARGE Quark structure of the pion and pion form factor}\\
{\Large V.Anisovich\footnote{email: anisovic@lnpi.spb.ru},
D.Melikhov
\footnote{Also at {\it Nuclear Physics Institute, Moscow State University},
email: melikhov@npi.msu.su},
and V.Nikonov\\
{\normalsize\it St.Petersburg Institute of Nuclear Physics, Gatchina, 188350,
Russia}
}
\end{center}

\vspace{2cm}
We consider the pion structure in the region of low and moderately high
momentum transfers: at low $Q^2$, the pion is treated as a composite system of
constituent quarks; at moderately high momentum transfers,
$Q^2=10\div25\;GeV^2$, the pion \ff is calculated within perturbative QCD
taking into account one--gluon hard exchange. Using the data on pion \ff at
$Q^2<3\;GeV^2$ and pion axial--vector decay constant, we reconstruct the pion
\wf in the soft and intermediate regions. This very wave function combined with
one--gluon hard scattering amplitude allows a calculation of the pion \ff in
the hard region $Q^2=10\div25\;GeV^2$. A specific feature of the reconstructed
pion \wf is a quasi--zone character of the $q\bar q$--excitations. On the basis
of the obtained pion \wf and the data on deep inelastic scattering off the
pion, the valence quark distribution in a constituent quark is determined.

\section{Introduction}
\begin{figure}
\begin{center}
\mbox{\epsfig{file=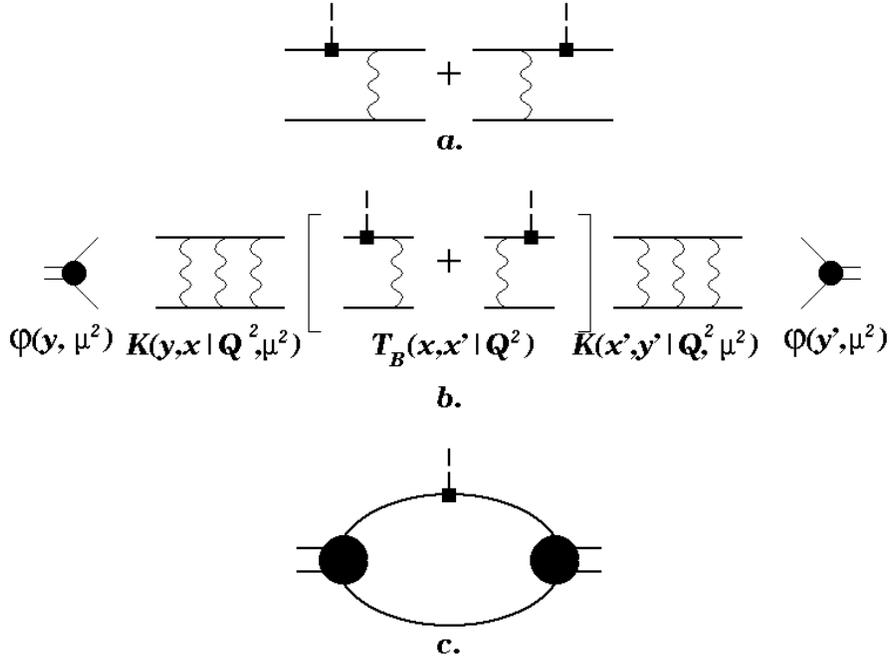,width=12cm}}
\end{center}
\caption{a. Hard scattering amplitude for two free quarks
in the Born approximation.
b. Hard scattering process in the LLA.
c. Soft form factor.
\label{fig:i1}}
\end{figure}

Perturbative QCD gives rigorous predictions for exclusive amplitudes,
in particular form factors at asymptotically large values of $Q^2$ \cite{hsp}.
For the pion \ff defined as
\beq
\label{ffdef}
<\pi(P')|J_\beta|\pi(P)>=(P'+P)_\beta F_\pi(Q^2)
\eeq
the pQCD result takes the form
\beq
\label{series}
F_\pi(Q^2)=8\pi f^2_\pi\frac{\alpha_s(Q^2)}{Q^2}
\left({
1+\sum\limits_{N=2,4,...}C_N \left({
{\rm ln}\frac{Q^2}{\mu^2}
}\right)^{-\gamma_N}
}\right)
+O(1/Q^4).
\eeq
Here $\gamma_N$ are the known positive numbers calculated within pQCD,
$\mu$ is a constant about 1 $GeV$ dividing the perturbative and
nonperturbative regions;
$f_\pi=130\;MeV$ is the pion axial--vector decay constant,
and the $C_N$ are expressed through the soft-region nonperturbative wave
function of the pion.
In the series (\ref{series}),
the contribution of diagrams with internal lines having virtualities
above $\mu$ are taken into account perturbatively, while all the exchanges
with lower virtualities are absorbed into the set of soft nonperturbative
wave functions of the pion Fock components.
The terms of the order $1/Q^2$
come only from the valence quark--antiquark component of the pion
Fock state, whereas other components give the terms $O(1/Q^4)$.
The series (\ref{series}) involves both the leading and subleading
logarithms. In the leading logaithmic approximation (LLA) the
expression (\ref{series}) can be rewritten in the form
\beq
\label{fflla}
F_\pi(Q^2)=\int_0^1 dx\,dx'\phi_\pi(x,Q^2)T_B(x,x',Q^2)\phi_\pi(x',Q^2)
\eeq
where $\phi(x,Q^2)$ is the leading twist wave function (distribution
amplitude) which describes the longitudinal momentum distribution
of valence quark--antiquark pair whose relative transverse momentum
is less than $Q$, and

\beq
\label{tbass}
T_B(x,x',Q^2)\to\frac{8\pi}9\frac{\alpha_s(xx'Q^2)}{xx'Q^2}
\eeq
is the amplitude of the hard interaction of the two free quarks
in the Born approximation (Fig.\ref{fig:i1}a).

The distribution amplitude at large $Q$ is related to the soft pion
distribution amplitude
$\phi_\pi(x,\mu^2)$ by the gluon ladder evolution kernel $K(x,y,Q^2/\mu^2)$
as follows
\beq
\label{daevol}
\phi_\pi(x,Q^2)=\int K(x,y,Q^2/\mu^2)\phi_\pi(y,\mu^2)\,dy
\eeq
The soft distribution amplitude is connected with the pion
axial vector decay constant via the relation
\beq
\label{danorm}
\int\phi_\pi(x,\mu^2)\;dx=f_\pi,
\quad f_\pi=130\;MeV
\eeq
So, in the LLA hard scattering off the pion described by the expressions
(\ref{fflla}) and (\ref{daevol}) (see Fig.\ref{fig:i1}b)
has a clear physical interpretation
within the hard scattering picture \cite{pqcd}, namely:
The initial pion transforms into quark--antiquark pair with a small
relative transverse momentum $\le\mu$ and the longitudinal momentum
fractions $x$ and $1-x$, respectively: this stage is described by
$\phi(y,\mu^2)$. Next, the quarks are coming closer to each other to the
distances $1/Q$ and increase their virtualities via ladder gluon exchanges,
described by $K$.
Then, the hard interaction of almost free quarks with the external
current occurs ($T_B$), and inverse evolution to low virtualities ($K$) with
a subsequent pion formation ($\phi(y',\mu^2)$).
A rigorous QCD result is that any soft distribution amplitude evolves at
large $Q^2$ to the universal function
\beq
\label{asswf}
\phi^{as}_\pi(x)=\phi_\pi(x, Q^2\to\infty)=6 f_\pi\, x\,(1-x).
\eeq
Substituted into (\ref{fflla}), this function gives the leading term in
(\ref{series}).

Unfortunately, this beautiful picture does not work at momentum transfers
accessiblle to present--day experiments: the momentum transfers are not large
enough. Trying to adjust this perturbative calculations for moderately high
momentum transfers,
Chernyak and Zhitnitsky \cite{cz} assumed the logarithmic and power
corrections (including a purely soft contribution shown if Fig.\ref{fig:i1})
to the Born term
to be not essential, and the pion form factor to be described by
\beq
\label{ffborn}
F_\pi(Q^2)=\int_0^1
dx\,\phi_\pi(x,\mu^2) T_B(x,x',Q^2) dx'\,\phi_\pi(x',\mu^2)
\eeq
with $T_B$ still given by (\ref{tbass}), but some modified distribution
amplitude $\phi_\pi$.
Describing the available pion
\ff data by the formula (\ref{ffborn}), they came to
the soft distribution amplitude of the form
\beq
\label{dacz}
\phi^{cz}_\pi(x,\mu\approx 0.5 GeV)=30f_\pi x(1-x)(1-2x)^2
\eeq

However, arguments against such an approach were put forward by
Isgur and Llewellin Smith \cite{ils}.
The problem is that the wave function
of the form (\ref{dacz})
strongly emphasizes the end-point region of $x\approx 0$
which was estimated to give $70\div90\%$ of pion
\ff at $Q^2=10\;GeV^2$.
Recall that only the exchanges with
virtualities above $\mu^2$ were considered perturbatively, otherwise
the corresponding subprocesses were referred to the soft wave function.
In the end--point region at moderately high $Q^2$,
the gluon virtuality $xx'Q^2$ is not large enough to justify the
perturbative treatment. Hence, the end--point contribution
should be rather referred to the nonperturbative one.
The large contribution coming from small $x$ means that in fact the \wf with
strongly emphasized end points has picked up a good portion of the
nonperturbative contribution. So the latter turns out to be not small that
contradicts to the inital assumption.

The arguments of \cite{ils} were supported by
by recent applications of QCD sum rules \cite{mr},\cite{bh}:
the end-point contribution remains numerically
important at least up to $Q^2\approx10\;GeV^2$, although parametrically it is
suppressed by an extra power of $1/Q^2$.
The problem of a correct extraction of the end--point contribution to hard
scattering amplitudes was studied by Li and Sterman \cite{ls} and Radyushkin
\cite{rad}. Their results give additional arguments against the application of
the strategy of ref.\cite{cz} to hadron \ffs at intermediate momentum
transfers.

In addition, some problems are encountered when \wfs with an emphasized
end--point region are applied
to deep inelastic scattering. It was discussed \cite{huang}
that the valence quark contribution to the deep inelastic structure
function calculated with such distribution amplitudes exceeds
the available data at large $x$, whereas there should be a room for other
Fock components.

The attempts to describe the pion \ff in the region $Q^2=3\div10\;GeV^2$ taking
into account only the perturbative hard scattering mechanism were not
successful. The nonperturbative contribution in this region is obviously not
small. Our goal is to consider the pion \ff at moderately large $Q^2$ allowing
for both the perturbative and nonperturbative contributions starting with low
$Q^2$ and advancing to higher values.

Investigations of soft hadron processes in past decades have demonstrated
constituent quarks to be relevant objects for describing hadron structure
\cite{fritzsch}\cite{w}.

The nonperturbative contribution to the pion \ff was considered
within the framework of
the QCD-inspired constituent quark model (\cite{ik}\cite{dziemb}
\cite{jaus}\cite{schlumpf}\cite{grach}), the pion \ff being represented by
the diagram of Fig.\ref{fig:i1}c.

\begin{figure}
\begin{center} \mbox{\epsfig{file=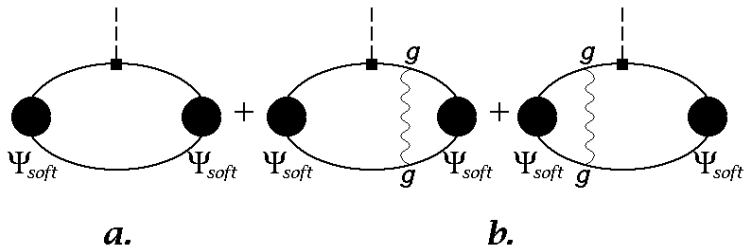,width=10cm}}
\end{center}
\caption{The expansion of the pion form factor  in the series over $g^2$.
\label{fig:i2}}
\end{figure}

In the approach suggested here, we move to the region of large $Q^2$ starting
with small values. That is, we take into account
the diagram of Fig.\ref{fig:i2}a and, as the following step, the diagrams of
Figs.\ref{fig:i2}b,c. The diagrams of Fig.\ref{fig:i2}b and Fig.\ref{fig:i2}c
correspond to the terms with the minimal $N=0$ in the series (\ref{series}).
In fact, we recast the series (\ref{series}) for the pion \ff as a series over
$\alpha_s$. Such an expansion is relevant at moderately large $Q^2$.

To be more quantitative, the procedure is as follows. The expansion for the
pion \ff in a series over $\alpha_s$ reads
\beq
\label{curdec}
J^\beta(0)=
[\bar q(0)\gamma^\beta q(0)]_{\mu^2}+
\alpha_s \int dz_1 dz_2 [\bar q(0)q(z_1)]_{\mu^2}|0>
T^\beta_B(z_1,z_2) <0|[\bar q(z_2)q(0)]_{\mu^2} +O(\alpha^2_s)
\eeq
To make this expansion meaningfull, the operator product expansion was
performed, i.e. all the field operators were decomposed into soft and hard
components.
The subscript $\mu^2$ implies that the subdiagrams for corresponding operators
involve only lines with virtualities below $\mu^2$.
The contribution of the region of virtualities below $\mu^2$ is described by
the
soft wave functions, whereas the contribution of larger virtualities is
represented by the hard scattering block $T_B$ \cite{hsp}.

Let us denote the corresponding contributions of the first and second terms in
the r.h.s. of (\ref{curdec}) as $F^{ss}$ (Fig.\ref{fig:i2}a) and $2F^{sh}$
(Figs.\ref{fig:i2}b,c), respectively. Then one finds
\beq
\label{ffdec}
F_\pi=F^{ss}+2F^{sh}+O(\alpha^2_s).
\eeq
The last series actually corresponds to dividing the pion
light--cone wave function $\Psi$ into two parts
such that $\Psi_{soft}$
is large at $s=\frac{m_\perp^2}{x(1-x)}<s_0$
while $\Psi_{hard}$ prevails at $s>s_0$.
We perform such a decomposition of the wave function using the simplest
ansatz with the step--function:
\beq
\label{psidec}
\Psi = \Psi_{soft}\theta(s_0-s) + \Psi_{hard} \theta(s-s_0)
\eeq
According to (\ref{curdec}),
$\Psi_{hard}$ is represented as a convolution of the one-gluon exchange
kernel $V^{\alpha_s}$ with $\Psi_{soft}$
\beq
\label{psihard}
\Psi_{hard}=V^{\alpha_s}\otimes \Psi_{soft}.
\eeq

The soft--soft contribution $F^{ss}$ in (\ref{ffdec})
is the usual quantity calculated within
constituent quark models, whereas the soft--hard term $F^{sh}$
relates to the one given by the hard scattering mechanism.
The soft--soft contribution includes the Sudakov form factor of the quark.\\
We obtain the following results:\\
(i) The pion \ff calculated in the region of $Q^2$ from 0 to 20 $GeV^2$
describes well the available data (Fig.\ref{graph:ff}).
The soft--soft contribution is found to give more than a half of the
\ff in the region $Q^2\le 20\; GeV^2$ and is not negligible up to
$Q^2\le 30\; GeV^2$.
However, the particular numbers depend on how we define the boundary of
the soft and the hard regions. We assume an extended soft region
for $s\le9\;GeV^2$, that yields a large contribution of the soft--soft
\ff until very high momentum transfers. \\
The transverse motion in the soft--hard term turns out to be
important. At the same time, the Sudakov suppression is not large in the
kinematical region of momentum transfers where the soft--soft term
dominates.\\
(ii) The soft pion wave function which has been a variational quantity
of our consideration is found to have a quasi--zone structure: it is large at
low $s\le2\;GeV^2$, then it almost vanishes at $2\;GeV^2\le s\le 4.5\;GeV^2$,
and has a bump at $4.5\;GeV^2\le s\le 9\;GeV^2$ (Fig.\ref{graph:wf}).\\
(iii) The pion structure function is expressed through pion soft \wf and
constituent quark structure function. By describing the data on pion
valence quark $x$--distribution, we find the parameters of the
$x$--distribution of valence quark inside a constituent quark to be in
a qualitative agreement with Reggeized QCD--gluon intercept calculation
\cite{bklf}.

The paper is organized as follows: \\
The Section 2 considers the pion as quark-antiquark bound state within the
light--cone technique \cite{terent} reformulated as
dispersion relation integrals and presents the expressions for the pion \ff and
and quark distribution in deep inelastic scattering.
All necessary technical details
relevant to pion description are given in the Appendices.
The results are discussed in the Section 3. A brief summary and outlook are
given in the Conclusion.

\section{Pion form factor and structure function}
The light--cone technique expressed in the form of the deispersion relation
integrals \cite{ammp} allows constructing relativistic and
gauge invariant amplitude of the interaction of a composite system
with an external vector field
starting with low-energy constituent scattering amplitude (see the Appendix A).
Two-particle $s$-channel interactions are consistently taken into account both
in the
constituent scattering amplitude and the amplitude of interaction with an
external
field. In the case of a bound state, its
\ff and structure function are expressed through \ff and structure function
of mass-shell constituents and the vertex $G$ of constituent--bound state
transition.
This vertex is defined by the two--particle irreducible block
of the constituent scattering amplitude.
On the one hand, the dispersion integral representation turns out to be
equivalent to the Bethe--Salpeter treatment with a separable kernel of
a special form, the vertex $G$ being
connected with the amputated Bethe-Salpeter wave function of the bound state.
On the other hand, this approach can be formulated as a light--cone description
of a bound state with the special form of spin transformation
(the Melosh rotation).
Because of the relativistic invariance, the dispersion integral
approach does not face the problem
of choosing appropriate component of the current for \ff calculation.
The function $G$
determines the bound--state light--cone wave function \cite{ammp}.
Note also that only amplitudes for on-shell constituents contribute to
corresponding
amplitudes of the bound state. This guarantees gauge invariance of the derived
expressions and escapes the problem of constituent amplitudes off the mass
shell. All relevant details can be found in the Appendix A.

Our position on the pion structure completely coincides with the viewpoint
formulated by Weinberg \cite{w}:
Successes of the quark model allow one to treat quarks as usual massive
hadrons with the only difference that quarks are subject to color forces which
become essential at large distances and keep quarks confined in hadrons;
in all other aspects these forces are weak  and quarks can be treated
as real particles.

In the soft region, quark structure of a pion is described by the vertex
\beq
\frac{\bar Q^a(P-k) i\gamma_5 Q^a(k)}{\sqrt{N_c}}G_v(P^2)
\eeq
with
$a$ a color index, $N_c=3$ the number of quark colors,
$k^2=m^2$, $(P-k)^2=m^2$, but $P^2=s\ne m_\pi^2$.
Here $m$ is a constituent quark mass.
We consider $\pi^+$ and omit the flavor which gives the unity factor.
(the appendix B presents a detailed consideration)
The soft vertex $G_v$ is supposed to be nonzero at $s<s_0$ in accordance
with (\ref{psidec}).
Once the vertex is fixed, we can proceed with \ff calculation.

\subsection{Soft--soft contribution to pion form factor}
The double dispersion relation integral for a soft--soft contribution to the
pion \ff is given by the following expression
\beq
\label{ffpi1}
F^{ss}(q^2)=\int\frac{ds\;G_v(s)}{\pi(s -M^2)}
\frac{ds' G_v(s')}{\pi(s'-M^2)}\Delta_\pi(s',s,q^2)f_c(q^2).
\eeq
Here $f_c$ is a constituent form factor, $f_c(0)=1$. The quantity
$\Delta_\pi$ is defined as follows
$$
-\frac1{8\pi}\int dk_1
dk'_1 dk_2 \delta(k^2_1-m^2)\delta(k'^2_1-m^2)\delta(k^2_2-m^2)
\delta(P-k_1-k_2)\delta(P'-k'_1-k_2)
$$
\beq
\label{deltapi1}
\times Sp\left({ (\hat k'_1+m)\gamma_\mu(\hat k_1+m)
i\gamma_5(m-\hat k_2)i\gamma_5 }\right)=2P_\mu(q) \Delta_\pi(s',s,q^2)
\eeq
with
$$
P_\mu(q)=(P-\frac{qP}{q^2}\,q)_\mu,\;P^2=s,\;P'^2=s',\;(P'-P)^2=q^2.
$$
The trace reads
\beq
\label{trace11}
\frac14 Sp\left({ (\hat k'_1+m)\gamma_\mu(\hat k_1+m)
\gamma_5(m-\hat k_2)\gamma_5 }\right)=
m^2(k_1+k'_1)_\mu+k_{1\mu}(k'_1k_2)+k'_{1\mu}(k_1k_2)-k_{2\mu}(qk_1)
\eeq
To reveal the relationship between the dispersion integral (\ref{ffpi1})
and the lihgt--cone technique, we introduce the light--cone variables
\beq
\label{lcvariables}
k_-=\frac{1}{\sqrt{2}}(k_0-k_z);\quad
k_+=\frac{1}{\sqrt{2}}(k_0+k_z);\quad
k^2=2k_+k_--k^2_\perp;
\eeq
into the representation (\ref{deltapi1}) and use the reference frame
\beq
\label{frame}
q_+=0,\qquad P_\perp=0,\qquad q^2=-q^2_\perp<0,
\eeq
The \ff takes the form
\beq\label{ffpilc1}
F^{ss}(q^2_\perp)=\frac1{\pi}\int dx d^2k_\perp
\psi(x,k_\perp)\psi(x,k_\perp-xq_\perp)\beta(x,k_\perp,q_\perp)
f_c(q^2_\perp),
\eeq
where the soft radial light-cone wave function of a pion is introduced
\beq
\label{wfpilc1}
\psi(x,k_\perp)=\frac{G_v(s)\sqrt{s}}{\pi\sqrt{8}(s-\mpis)\sqrt{x(1-x)}},
\qquad s=\frac{m^2+k^2_\perp}{x(1-x)}
\eeq
$$
\beta=\frac{m^2+k^2_\perp-xk_\perp q_\perp}{\sqrt{m^2+k^2_\perp}\;
\sqrt{ m^2+(k_\perp-xq_\perp)^2}}.
$$
The quantity $\beta$ accounts for the contribution of spins.
It is different from unity at $q_\perp\ne 0$
because both the nonspin-flip and
spin-flip amplitudes of the interacting quark contribute.
The relation $F_\pi(0)=1$ is the normalization condition
for the soft radial wave function
\beq
\int dx\;dk^2_\perp\;|\psi(x,k_\perp)|^2=1.
\eeq
The pion axial--vector decay constant $f_\pi$ related to the $\pi\to\mu\nu$
decay is given by
\beq
\label{fpilc}
g_A\frac{\sqrt{N_c}}{\sqrt2 \pi}
\int{dx\;dk^2_\perp}\frac{m}{\sqrt{m^2+k^2_\perp}}\;\psi(x,k_\perp)={f_\pi}
\eeq
where $g_A$ is the constituent quark axial--vector coupling
constant.
{}From the analysis of the neutron $\beta$--decay within
various models, the value of $g_A$ was found to be in the range from
0.75 (nonrelativistic constituent quark model without configuration
mixing in the nucleon wave function, $m\approx 0.33\;GeV$) up to 1.0
(relativistic quark model with light constituent quarks,
$m\approx 0.25\;GeV$ \cite{schlumpf}.)
Here we use the relation (\ref{fpilc}) to fix the value of $g_A$
related to a particular pion wave function
such that (\ref{fpilc}) reproduces the observed value $f_\pi=130\;MeV$.
The value of $g_A$ is found to lie in the range $g_A\approx 0.75\div 1$
(see the Table \ref{table:par}).

The same expressions for the \ff and pion electroweak constant as
(\ref{ffpilc1})--(\ref{fpilc}) were derived in refs
\cite{jaus}--\cite{grach}. In contrast to the mentioned
papers where such formulas were applied to describing the total pion \ff,
we use them for calculating the soft--soft contribution.

The soft-soft \ff involves the constituent quark \ff which
should satisfy the conditions $f_c(0)=1$ and $f_c(Q^2)\to S(Q^2)$ at
large $Q^2=-q^2$. Here $S(Q^2)$ is the Sudakov \ff which
is taken in the form \cite{rad}
\beq
S(Q^2)={\rm exp}\({-\frac{\alpha_s(Q^2)}{2\pi}C_F
{\rm log}^2 \( {\frac{Q^2}{Q^2_0}} \) }\),
\; C_F=\frac{N_c^2-1}{2N_c}=\frac43,\; Q_0\approx0.85\div 1\;GeV
\eeq
with $\alpha_s$ the coupling constant.
At low $Q^2$ we assume $\alpha_s$ to be frozen at $1\; GeV^2$,
namely we set
\beeq
\label{alpha}
\nonumber
\alpha_s(Q^2)&=&\frac{4\pi}9{\rm log}^{-1}\({ \frac{Q^2}{\Lambda^2} }\),
\;\Lambda=0.22\;GeV, \quad Q^2<1\;GeV^2
 \\
&=&{\rm const,}\quad Q^2<1\;GeV^2
\eeeq
The constituent quark \ff is taken as
\beeq
f_c(Q^2)&=&1,\;Q<Q_0      \\
&=&S(Q^2),\;Q>Q_0,\;S(Q_0^2)=1 \nonumber
\eeeq
Note that the Sudakov suppression is absent in
the soft-hard term.

\subsection{Soft--hard contribution to the pion form factor}
In accordance with (\ref{curdec}), the soft--hard contribution
is described by the two graphs of Fig.\ref{fig:i2}b,c with one--gluon
exchanges.
The corresponding dispersion relation integral reads
\beq
\label{ffmod}
F^{sh}=\int\frac{ds\;G_v(s)\;\theta(s<s_0)}{\pi(s -\mpis)}
\frac{ds''\theta(s''>s_0)}{\pi(s''-\mpis)}D(s,s'',s',q^2)
\frac{ds'\;G_v(s')\;\theta(s'<s_0)}{\pi(s' -\mpis)}
\eeq
In this expression, $D(s,s'',s',q^2)$ is a spectral density
which takes into account the one-gluon exchange at large $s''$;
the Appendix C presents the details of the soft--hard \ff $F^{sh}$ calculation.
Notice that the dispersion expression involves only two--particle
singularities and neglects three--particle intermediate state related
to cutting the gluon line.
The final result has the form
\beq
\label{shff1}
F^{sh}(q_\perp^2)=\frac{C_F}{8\pi}
\int
\frac{dxd^2k_\perp}{\pi\sqrt{m^2+k^2_\perp}(1-x)}\psi(x,k_\perp)\theta(s<s_0)
\frac{\theta(s''>s_0)}{\frac{m^2+(k_\perp-xq_\perp)^2}{x(1-x)}-\mpis}
\eeq
$$
\times\frac{dx'd^2k'_\perp}{\sqrt{m^2+(k'_\perp-x'q_\perp)^2}}
\psi(x',k'_\perp-x'q_\perp)\theta(s'<s_0)
Tr(s,s',s'',q^2)\frac{\alpha_s\({-(k-k')^2}\)}{m_G^2-(k'-k)^2}
$$
where
$$
-(k'-k)^2=\frac{(k'_\perp x-k_\perp x')^2+m^2(x-x')^2}{xx'},
$$
$$
Tr(s,s',s'',q^2)=2s'\left({\alpha(s'',s,q^2)(s''+s-q^2)+q^2}\right)
-4m^2\left({\alpha(s'',s,q^2)(s'+s-q^2)+q^2)}\right);
$$
$$
\alpha(s'',s,q^2)=\frac{-q^2(s''+s-q^2)}{(s''-s)^2-2q^2(s''+s)+q^4}
$$
and
$$
s=\frac{m^2+k_\perp^2}{x(1-x)},\quad
s''=\frac{m^2+(k_\perp-xq_\perp)^2}{x(1-x)},\quad
s'=\frac{m^2+(k'_\perp-x'q_\perp)^2}{x'(1-x')}
$$
In (\ref{shff1})
$m_G$ is the gluon mass which depends on the
gluon momentum squared $-(k'-k)^2$
and is normalized to be of the order of $1\;GeV$ in the soft region
\cite{parisi}. \\
One can easily see that $F^{sh}(0)=0$. \\
Let us turn to the region of
large $Q^2$ and compare
the soft--hard term which dominates the form factor with the standard pQCD
expression (\ref{fflla}) and (\ref{tbass}).
Using the relations
$$
Tr\to 4m^2Q^2x \qquad{\rm and}\qquad -(k'_2-k_2)^2\to xx'Q^2
$$
we find
\beq
\label{ffpidass}
F_\pi(Q^2)\to\frac{1}{\kappa^2}
\int\;\phi(x,s_0)\;dx
\frac{8\pi}9\frac{\alpha_s(xx'Q^2)}{xx'Q^2}
\phi(x',s_0)\;dx'
\eeq
Here we introduced the distribution amplitude $\phi(x,s_0)$
normalized by the standard condition
\beq
\int\phi(x,s_0)\;dx=f_\pi
\eeq
This distribution amplitude is related to the soft radial wave function
(\ref{wfpilc1}) via the
relation
\beq
\label{daphi}
\phi(x, s_0)=\frac{\sqrt{N_c}\kappa}{\sqrt2\pi}
\int dk_\perp^2 \frac{m}{\sqrt{m^2+k_\perp^2}}\psi(x,k_\perp)
\theta\({\frac{m^2+k_\perp^2}{x(1-x)}<s_0}\)
\eeq
The expression (\ref{wfpilc1}) differs from (\ref{ffborn})
by the factor $1/\kappa^2$, where
\beq
\kappa=g^0_A(0)=0,75\div1
\eeq
(see the eq.(\ref{kappa}) of the Appendix B) is the axial--vector decay
constant at the level of a constituent quark. If we identify constituent
quarks with a bare pQCD quarks, then $g^0_A=1$. Here we assume such an
identification to be not well--justified in the region $Q^2\le 10\;GeV^2$ and
hence we allow $\kappa$ to be in the range from 0.75 to 1.

Actually, a constituent quark is a Fock state involving the components
$q_{val}, q_{val}\bar qq, q_{val}\bar qqg,\ldots$ At asymptotically large
momentum transfers only the valence component survives whereas all other
components do not contribute to the form factor. The problem of the typical
$Q$ at which  the transfer to the asymptotic regime occurs is still opened.
It is quite natural that the expression (\ref{ffpidass}) gives a larger
value for the \ff at $\kappa<1$ since it includes the contribution of all
higher
Fock components of the constituent quark in addition to the valence one.

\subsection{Structure function}
Now we have all necessary to calculate the pion structure function $F_2(x)$.
This function is given by the convolution of the pion light--cone soft
wave function
and the corresponding constituent quark structure function $f^i_2(x)$
\beq
F_2(x,Q^2)=\sum\limits_{j=Q,\bar Q}
\int_x^1 dk^2_\perp dx' \psi(x',k_\perp)^2
f^j_2\({\frac{x}{x'},Q^2}\)
\eeq
The leading order expressions for these structure functions
in terms of parton densities read
\beq
F_2(x)=\sum\limits_{flavours}e_i^2 x\;Q_i(x),\quad
f^j_2(x)=\sum\limits_{flavours}e_i^2 x\;q^j_i(x)
\eeq
where $Q_i$ and $q^j_i$ are quark parton densities of a given flavour $i$
inside a pion and constituent quark $j$, respectively.

At large $x$ the behavior of the structure functions is determined by
the valence quark (antiquark)
distribution $v(x)$ inside the constituent quark (antiquark).
The latter is taken in the form
\beq
\label{valence}
v(x)\approx x^{-a}(1-x)^{-b}\({1-N(1-x)^{c}}\)
\eeq
The parameters in this formula are chosen such that $v(x)$
should satisfy the following sum rules at $Q_0^2=10\;GeV^2$
\beq
\int\limits_0^1 v(x,Q_0^2)dx=1\qquad \int\limits_0^1 v(x,Q_0^2)x dx=0.35
\eeq
The last number is the fraction of constituent quark momentum carried by the
valence
quark. Due to the sum rule for momentum, this value is equal to the
fraction of hadron momentum carried by valence quarks. For a proton this
fraction is known to be $\approx0.35$ at $10\;GeV^2$.

\section{The results of fitting the data on \ff and deep inelastic parton
distribution}

\begin{figure}
\begin{center}
\vspace{-2cm}
\mbox{\epsfig{file=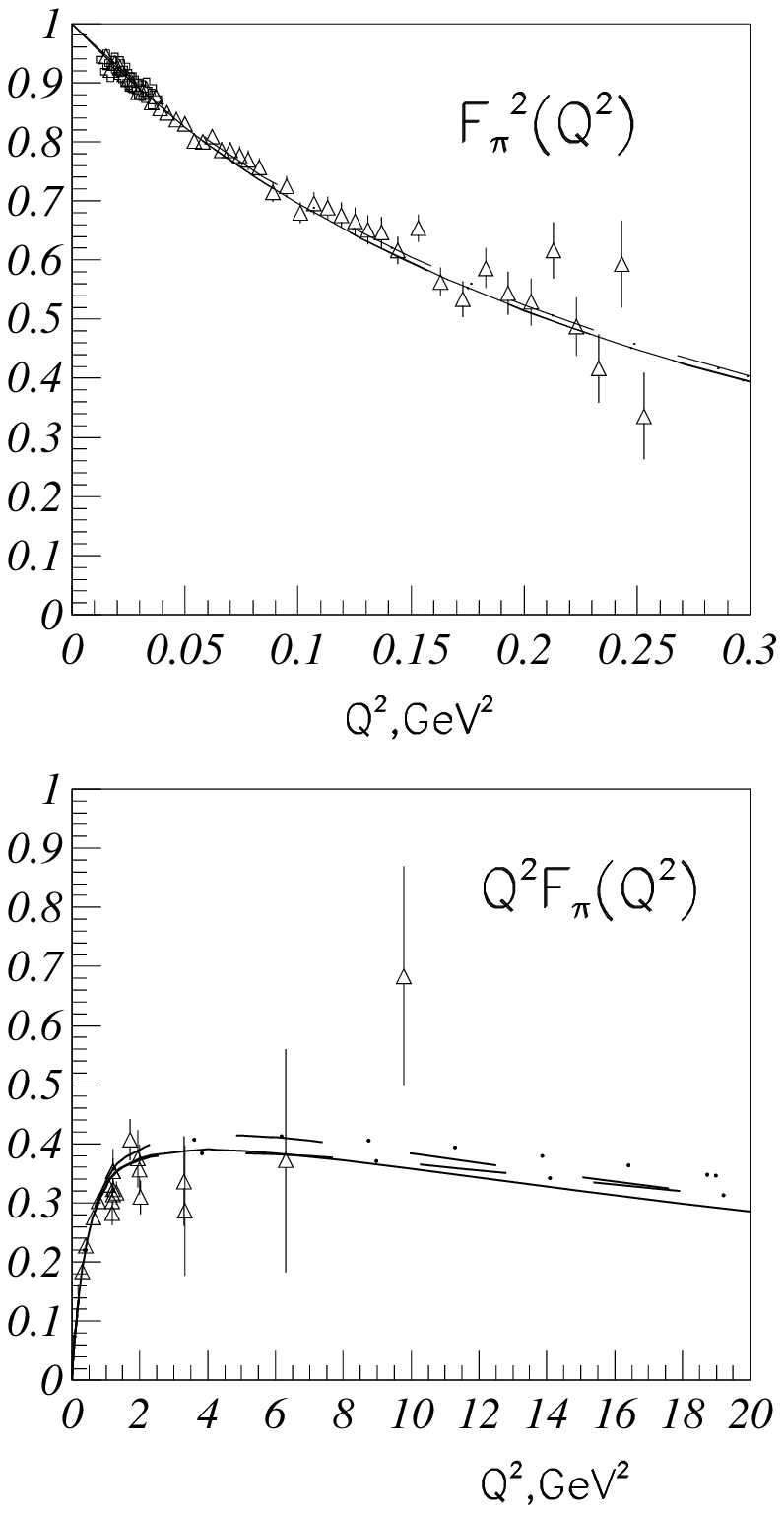,height=22cm}}
\end{center}
\caption{The pion form factor. Solid line -- calculations with the Set 1 of the
parameters;
dashed--dotted line -- Set 2, dashed line -- Set 3, dotted line -- Set 4.
\label{graph:ff}}
\end{figure}

\begin{figure}
\begin{center} \mbox{\epsfig{file=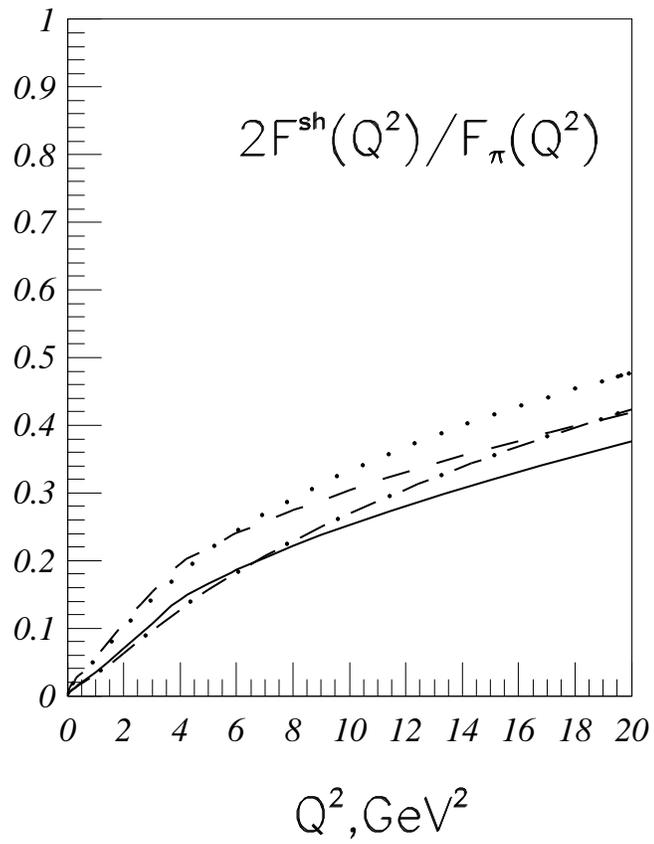,height=15cm}}
\end{center}
\caption{The contribution of the soft--hard term to the pion form factor.
The curve notation is the same as for the form factor.
\label{graph:ratio}}
\end{figure}

Fig.\ref{graph:ff} shows the results of fitting the data on $F_\pi(Q^2)$
\cite{bebek} via the
formulas (\ref{ffpilc1}) and (\ref{wfpilc1}) for $F^{ss}$ and (\ref{shff1})
for $F^{sh}$. Fig.\ref{graph:ff}a demonstrates the region of low $Q^2$ where
$F^{ss}$ dominates; Fig.\ref{graph:ff}b emphasizes the region of moderately
large
$Q^2$, $Q^2=5\div20\;GeV^2$.
Fig.\ref{graph:ratio} plots the typical contribution of the soft--hard term
to the pion form factor.
Several variants of the calculation relate to
different sets of the parameters $m_G, \alpha_s$, and $m$
presented in the Table \ref{table:par}.
Let us point out that $g_A^0$ calculated with the
determined \wf appreciably depends on the constituent quark mass: for a heavy
constituent quark $g^0_A$ is rather small and equal to 0.86, whereas for a
light constituent quark $g^0_A$  is close to unity.

\begin{figure}
\begin{center}
\vspace{-2cm}
\mbox{\epsfig{file=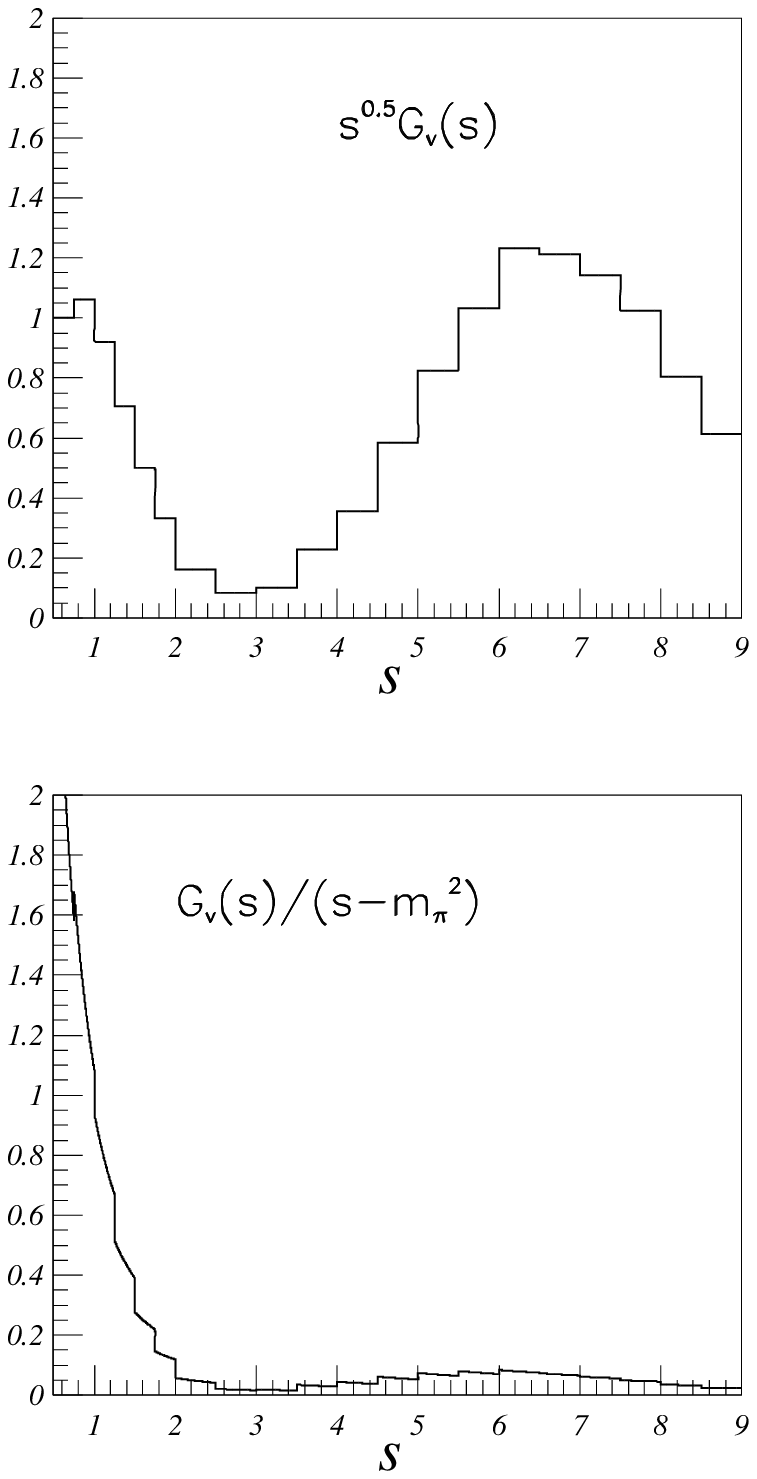,height=22cm}}
\end{center}
\caption{The soft light--cone \wf of the pion.
\label{graph:wf}}
\end{figure}

In all the fits the soft pion \wf $\psi(x,k^2_\perp)$ was the basic variational
quantity. Fig.\ref{graph:wf} presents the typical reconstructed soft
wave function.
A specific feature of the reconstructed \wf is a quasi--zone
character of $\bar qq$--excitations in the pion. We tried to find a
parametrization without a dip in the region $2\div4\;GeV^2$ but failed;
all the used fitting procedures suggested qualitatively similar
double--humped wave functions.
So one can think this dip to reflect some essential feature of the $\bar qq$
dynamics inside the pion. However, we have no definite ideas on the origin of
such a specific behavior.

\begin{figure}
\begin{center} \mbox{\epsfig{file=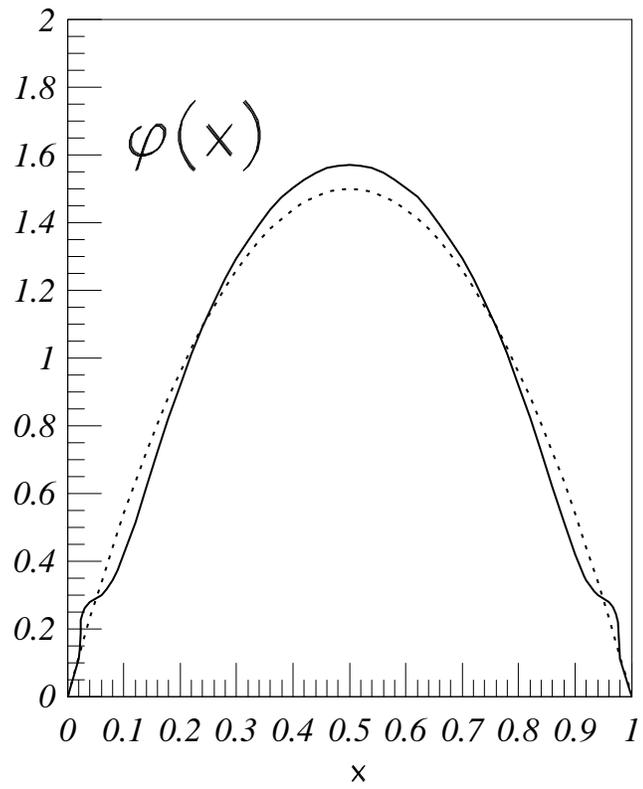,height=15cm}}
\end{center}
\caption{The distribution amplitude. Solid line -- our calculation;
dashed line -- asymptotical curve $\phi(x)=6x(1-x)$.
\label{graph:da}}
\end{figure}

Fig.\ref{graph:da} displays the distribution amplitude $\phi(x)$ calculated
with the determined $G_v(s)$. It turned out to be very close to the asymptotic
function (\ref{asswf}).

\begin{figure}
\begin{center} \mbox{\epsfig{file=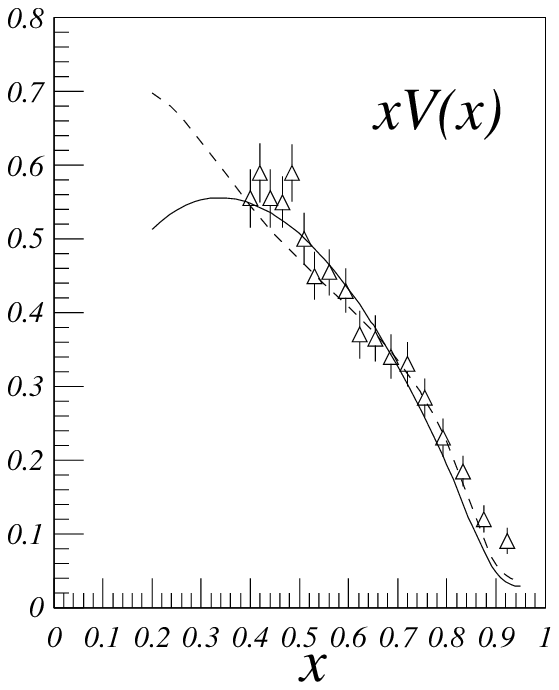,height=15cm}}
\end{center}
\caption{Valence distribution in a pion $x\,V(x, Q^2=10\; GeV^2)$.
The solid curve -- $a=0.69,\;b=0.99,\;c=1$;
the dashed curve -- $a=0.46,\;b=0.99,\;c=1.55$.
\label{graph:sf}}
\end{figure}

Fig.\ref{graph:sf} presents the valence quark distribution inside the pion.
The parameters of the distribution (\ref{valence}) $b=0.99$ and $a$ and $c$
from the range $a=0.5\div 0.7$ and $c=1\div2$
were found to provide a reasonable description of the data \cite{pionsf}.
Mention that $b$ is very close to unity. If the valence
quark distribution is determined by the Reggeized--gluon exchange, then our
result gives for its intercept the value close to unity in qualitative
agreement with \cite{bklf}.

\begin{table}
\caption{\label{table:par}
Table of the parameters in the pion \ff calculations.
We assumed $m_G(\kappa^2)=m_G^0(1-\kappa^2/0.5)$ as $\kappa^2<0.5\;GeV^2$ and
$m_G=0$ as $\kappa^2>0.5\;GeV^2$, where $\kappa$ is the gluon momentum. }
\centering
\begin{tabular}{|c|c|c|c|c|}
\hline
   & Set 1   & Set 2 & Set 3  & Set4 \\
 \hline\hline
constituent quark mass $m$, GeV & 0.35 & 0.35 & 0.35 & 0.35 \\
\hline
$m_G^0$, GeV
& 0.7 &  0.7 & 0&0 \\
\hline
$\alpha_s$ in (\ref{shff1})& $\alpha_s(Q^2/4)$  & $\alpha_s(\kappa^2)$
& $\alpha_s(Q^2/4)$& $\alpha_s(\kappa^2)$ \\
\hline
$g_A(0)$ & 0.86  & 0.86 & 0.86 &0.86 \\
\hline\hline
\end{tabular}
\end{table}

\section{Conclusion}
We have analyzed the pion structure within a dispersion
relation formulation of the light--cone technique when the pion in the soft
region is treated as a two constituent quark bound state.
At small values of the $\bar qq$ light--cone energy the pion
is described by a model soft wave function, whereas at large
values the one--gluon hard exchange is taken into account.
This provides the \ff high--$Q^2$ asymptotic behavior in agreement with pQCD.
The obtained pion \ff describes well the available experimental data at
$Q^2\div10\;GeV^2$. \\
Our results are as follows:
\begin{itemize}
\item
We considered the soft--soft ($F^{ss}$) and soft--hard ($2F^{sh}$)
contributions
to the pion \ff within the light--cone quantum mechanics. The derived
expressions involve the soft radial \wf of the pion which has been treated as
a variational parameter of the approach. By fitting the data on the pion \ff
at $Q^2\le3\;GeV^2$, we determined this soft radial wave function.
This allowed a calculation of the relative soft--hard contribution to the
pion \ff in a broad range of momentum transfers. It turned out to be
relatively small (less than 50\% at $Q^2=20\;GeV^2$) because we used a large
value for the boundary of light--cone energy squared dividing the soft and
the hard regions ($s_0=9\;GeV^2$) and hence we related a large portion of the
pion form factor to the soft--soft contribution. However, smaller values of
this boundary do not change qualitatively the results, except for
quantitative increasing the soft--hard fraction. \\
The calculated pion axial--vector decay constant agrees well
with the experimental value.

\item
The soft radial light--cone \wf as a function of the square of the light--cone
$\bar qq$
energy $s$ has been found to demonstrate a specific behavior: it is large at
$s\le2\;GeV^2$, close to zero as $s=2\div4\;GeV^2$, and has a bump in the
region $s=4\div9\;GeV^2$. Our attempts to find a \wf of a more regular shape
failed as all the fits suggested a double--humped behavior.
We have no definite ideas about the origin of such a quasi--zone
character of the $\bar qq$ excitations. Nevertheless,
such an unexpected \wf seems to reflect some unknown essential details of the
quark dynamics in the pion.

\item
The distribution amplitude $\phi(x)$ calculated with the obtained soft \wf
turns out to be very close to the asymptotic function $\phi^{as}_\pi(x)$
predicted by pQCD.

\item
Describing the data on deep inelastic scattering off the pion allowed
investing the parton structure of the constituent quark. The distribution of
the valence quark--parton in the constituent quark is found to be in a
qualitative agreement with the parametrization suggested by the
Reggeized--gluon exchange.
\end{itemize}

The authors are grateful to the International Science Foundation for
financial support under grant R1000.

\section{Appendix A: Bound state description within dispersion relations}
\begin{figure}
\begin{center}  \mbox{   \epsfig{file=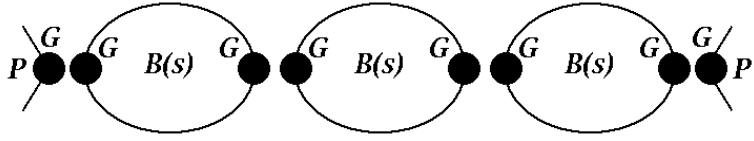,height=2cm}    }
\end{center}
\caption{ One of the terms in the expansion of $A_0(s)$
\label{fig:1}}
\end{figure}

To illustrate main points of the dispersion approach we
consider the case of two spinless constituents with the masses $m$ interacting
via
exchanges of a meson with the mass $\mu$. We start with the scattering
amplitude
\beq
A(s,t)=<k'_1,k'_2|S|k_1,k_2>, \quad s=(k_1+k_2)^2,\; t=(k_1-k'_1)^2
\eeq
The amplitude as a function of $s$ has the
threshold singularities in the complex $s$-plane connected with
elastic rescatterings of the constituents and production of new mesons at
\beq
s= 4m^2,\;(2m+\mu)^2,\;(2m+2\mu)^2\ldots
\eeq
We assume that an $S$-wave bound state with the mass $M<2m$ exists, then
the partial amplitude $A_0(s)$ has a pole at $s=M^2$.
The amplitude $A(s,t)$ has also $t$-channel singularities at
$t=(n\mu)^2;\;\; n=1,2,3\ldots$ connected with meson exchanges.
If one needs to construct the amplitude in the low-energy region $s\geq 4m^2$
the dispersion $N/D$ representation turns out to be convenient.
Consider the $S$-wave partial amplitude
\beq
A_0(s)=\int\limits^1_{-1} dz\, A(s,t(s,z)),
\eeq
where
$t(z)=-2(s/4\,-m^2)(1-z)$, $z=\cos\theta$ in the c.m.s.
The $A_0(s)$ as a function of complex $s$ has the right-hand
singularities related to $s$-channel singularities of $A(s,t)$.
In addition, it has left-hand singularities located at
$s=4m^2-(n\mu)^2;\;\; n=1,2,3\ldots$. They come from $t$-channel singularities
of $A(s,t)$.
The unitarity condition in the region $s\approx 4m^2$ reads
\beq
{\rm Im}\, A_0(s)= \rho(s)\;|A_0(s)|^2,\qquad
\rho(s) =\frac1{16\pi}\sqrt{1-\frac{4m^2}{s}}
\eeq
with
$\rho(s)$ the two-particle phase space.
The $N/D$ method represents the partial amplitude as $A_0(s)=N(s)/D(s)$, where
the function $N$ has only left-hand singularities and $D$ has only right-hand
ones. The unitarity condition yields
\beq
D(s) = 1 - \int\limits^\infty_{4m^2}\frac{d\ts}{\pi}\, \frac{\rho(\ts)N
(\ts)}{\ts - s}\; \equiv \; 1-B(s).
\eeq
Assuming the function $N$ to be positive we introduce
$G(s)=\sqrt{N(s)}$.
Then the partial amplitude takes the form
\beq
A_0(s)=G(s)\left[1+B(s)+B^2(s)+B^3(s)+\ldots\right] G(s)
=\frac{G(s)G(s)}{1-B(s)}.
\eeq
This expression can be interpreted as a series of loop diagrams of
Fig.\ref{fig:1}
with the basic loop diagram
\beq
B(s)=\int\limits^\infty_{4m^2}\frac{d\ts}{\pi}\, \frac{\rho(\ts)\;
G^2(\ts)}{\ts - s}.
\eeq
The bound state with the mass $M$ relates to a pole both in the total
and partial amplitudes at $s=M^2$ so $B(M^2)=1$.
Near the pole one has for the total amplitude
\beeq
A&=&<k'_1,k'_2|P>\frac1{M^2-P^2}<P|k_1,k_2>+{\rm regular\; terms} \nonumber \\
&\equiv&\chi^*_P(k'_1,k'_2)\frac1{M^2-P^2}\chi_P(k_1,k_2)+\ldots
\eeeq
where $\chi_P(k_1,k_2)$ is the amputated Bethe-Salpeter amplitude
of the bound state.
The dispersion amplitude near the pole reads
$$
A=N/D+{\rm regular\; terms\; related\; to\; other\; partial\; waves}
$$
\beq
=\frac{G^2(M^2)}{(M^2-s)B'(M^2)}+\ldots
\equiv \frac{G_v^2(M^2)}{M^2-s}+\ldots
\eeq
where $G_v$ is a vertex of the bound state transition to the constituents.
The singular terms correspond to each other and hence
\beq
\chi_P(k_1,k_2)\to G_v(P^2)\equiv \frac{G(P^2)}{\sqrt{B'(M^2)}} \label{bsa}
\eeq
Underline that among right-hand singularities the constructed dispersion
amplitude takes into account only the two-particle cut.

Let us turn to the interaction of the two-constituent system with an external
electromagnetic field. The amplitude of this process
$T_\mu=<k'_1,k'_2|J_\mu(q)|k_1,k_2>$ in the case of a bound state takes the
form
\beeq
T_\mu&=& <k'_1,k'_2|P'>
\frac1{P'^2-M^2}
<P'|J_\mu(q)|P>
\frac1{P^2-M^2}
<P|k_1,k_2>+\ldots  \label{tmu} \nonumber \\
&=&\chi^*_P(k'_1,k'_2)
\frac1{P'^2-M^2}
(P'+P)_\mu F(q^2)
\frac1{P^2-M^2}
\chi_P(k_1,k_2)+\ldots
\eeeq
where the bound state \ff is defined as
\beq
<P'|J_\mu(q)|P>=(P'+P)_\mu F(q^2)
\eeq

\begin{figure}
 \begin{center}  \mbox{
   \epsfig{file=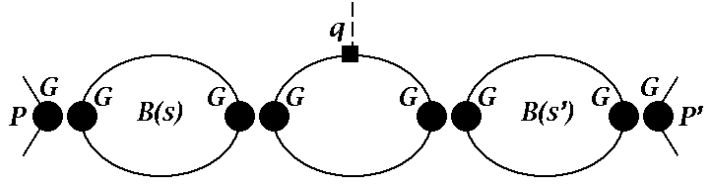,height=2.5cm}
                     }
\end{center}
\caption{ One of the terms in the series for $T_\mu$.
\label{fig:2}}
\end{figure}

The dispersion amplitude
$T_\mu$ with only two-particle singularities in the $P^2$- and
$P'^2$-channels taken into account is given \cite{akms} by the series of graphs
in Fig.\ref{fig:2}.

These graphs are obtained from the dispersion scattering
amplitude series by inserting a photon line into constituent lines.
The amplitude reads
\beq
T_\mu(P',P,q)=2P_\mu(q)T(s',s,q^2)+\frac{q_\mu}{q^2}C,
\eeq
$$
P^2=s,\;P'^2=s',\;q=P'-P,\;P_\mu(q)=(P-\frac{qP}{q^2}\,q)_\mu
$$
The dispersion method allows one to determine
$T(s,s',q^2)$, which is the part of the amplitude transverse with respect
to $q_\mu$.
Summing up the series of dispersion graphs in Fig.\ref{fig:2} gives
\beq
T(s',s,q^2)=\frac{G(s)}{1-B(s)}\Gamma(s',s,q^2)\frac{G(s')}{1-B(s')}.
\eeq
Here
$$
\Gamma(s',s,q^2)=\int\frac{d\ts G(\ts)}{\pi(\ts -s)}
\frac{d\ts ' G(\ts')}{\pi(\ts '-s)}\Delta(\ts',\ts,q^2),
$$
and $\Delta(\ts',\ts,q^2)$
is the double spectral density of the three-point Feynman graph with a
pointlike vertex of the constituent interaction.

The longitudinal part $C$ is given by the Ward identity
\beq
C=\frac{G(s)}{1-B(s)}\left( {B(s')-B(s)} \right) \frac{G(s')}{1-B(s')}
\eeq

At $s=s'=M^2$, the quantity $T_\mu$ develops both $s$ and $s'$ poles, so
\beq
\label{ff}
T_\mu(P',P,q)=\frac{G_v(M^2)}{M^2-s}(P'+P)_\mu F(q^2)
\frac{G_v(M^2)}{M^2-s'}+{\rm less\;singular\;terms}
\eeq
where
\beq
\label{ffv}
F(q^2)=\int\frac{ds G_v (s)}{\pi(s -M^2)}
\frac{ds' G_v (s')}{\pi(s'-M^2)}\Delta(s',s,q^2).
\eeq
is the bound--state \ff (see (\ref{bsa}) and (\ref{tmu})).
So, the quantity $<P'|J_\mu(q)|P>$ corresponds to the three--point dispersion
graph
with the vertices $G_v$.
The following relation is valid
$\Delta(s',s,0)=\pi\delta(s'-s)\rho(s)$.
This is a consequence of the Ward identity which relates the
three-point graph at zero
momentum transfer to the loop graph. This relation yields the charge
normalization $F(0)=1$. The expression (\ref{ffv}) gives the \ff in terms of
the $N$-function of the constituent scattering amplitude and double
spectral density of the Feynman graph.
In general, the following prescription works: to obtain the dispersion
expression
spectral density in channels corresponding to a bound state, one should
calculate
the related Feynman graph spectral density and multiply it by $G_v$.

Mention that
only on-shell constituents contribute to all the quantities.
If the constituent is a nonpoint particle,
the expression (\ref{ffv}) should be multiplied by \ff of an on-shell
constituent.

\section{Appendix B: Pion within the dispersion approach}
\begin{figure}
 \begin{center}  \mbox{
   \epsfig{file=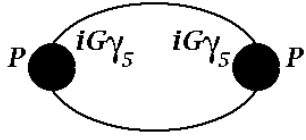,height=2.cm}
                     }
\end{center}
\caption{Pionic dispersion loop graph $B_\pi(P^2)$.\label{fig:3}}
\end{figure}

We start with describing the quark-pion
vertex. For on-shell quarks there is one independent structure
$i \bar q^a\gamma_5 q^a/\sqrt{N_c}$,
with $a$ a color index and $N_c=3$ the number of quark colors. We consider a
$\pi^+$ and omit the flavor which gives the unity factor.
Let us introduce the momentum $P_\pi$,
$P_\pi^2=m^2_\pi$.
We shall also use an off-shell pion momentum $P$, $P^2=s$.
The pionic dispersion loop graph Fig.\ref{fig:3} reads
\beq
B_\pi (P^2)=\int\limits^\infty_{4m^2}\frac{ds\;G^2(s)}{\pi(s-P^2)}\rho_\pi(s),
\quad B_\pi(m^2_\pi)=1
\eeq
with $\rho_\pi(s)$ the spectral density of the Feynman loop graph
\beq
\rho_\pi(s)=-\frac1{8\pi^2}\int dk_1 dk_2\delta(k^2_1-m^2)\delta(k^2_2-m^2)
\delta(P-k_1-k_2)\;
Sp\left({ (\hat k_1+m)i\gamma_5(m-\hat k_2)i\gamma_5 }\right)
\eeq
$$
=\frac{s}{8\pi}\sqrt{1-\frac{4m^2}{s}}\;\theta(s-4m^2)
$$
with $m$ a constituent quark mass.

\begin{figure}
 \begin{center}  \mbox{
   \epsfig{file=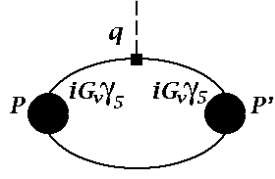,height=2.5cm}
                     }
\end{center}
\caption{The soft--soft contribution $F^{ss}_\pi(q^2)$ to pion form factor.
\label{fig:4}}
\end{figure}

The \ff of a pion is given by the following matrix element
\beq
<P'_\pi|J^{em}_\mu(0)|P_\pi>=(P'_\pi+P_\pi)_\mu\;F_\pi(q^2)
\eeq
$$P^2_\pi=P'^2_\pi=M^2,\;(P'_\pi-P_\pi)^2=q^2.
$$
The double dispersion representation for the \ff corresponds to
Fig.\ref{fig:4}
\beq
\label{ffpi}
F_\pi(q^2)=\int\frac{ds\;G_v(s)}{\pi(s -M^2)}
\frac{ds' G_v(s')}{\pi(s'-M^2)}\Delta_\pi(s',s,q^2)f_c(q^2),\quad
G_v(s)=\frac{G(s)}{ \sqrt{B'(\mpis)} }
\eeq
Here $f_c$ is a constituent form factor, $f_c(0)=1$, and
$\Delta_\pi$ is defined as
$$
-\frac1{8\pi}\int dk_1
dk'_1 dk_2 \delta(k^2_1-m^2)\delta(k'^2_1-m^2)\delta(k^2_2-m^2)
\delta(P-k_1-k_2)\delta(P'-k'_1-k_2)
$$
\beq
\label{deltapi}
\times Sp\left({ (\hat k'_1+m)\gamma_\mu(\hat k_1+m)
i\gamma_5(m-\hat k_2)i\gamma_5 }\right)=2P_\mu(q) \Delta_\pi(s',s,q^2)
\eeq
with
$$
P_\mu(q)=(P-\frac{qP}{q^2}\,q)_\mu,\;P^2=s,\;P'^2=s',\;(P'-P)^2=q^2.
$$
The trace reads
\beq
\label{trace1}
\frac14 Sp\left({ (\hat k'_1+m)\gamma_\mu(\hat k_1+m)
\gamma_5(m-\hat k_2)\gamma_5 }\right)=
m^2(k_1+k'_1)_\mu+k_{1\mu}(k'_1k_2)+k'_{1\mu}(k_1k_2)-k_{2\mu}(qk_1)
\eeq
Multiplying both sides of (\ref{deltapi}) by $P_\mu$ and using (\ref{trace1})
one obtains
\beq
\Delta_\pi(s',s,q^2)=\frac{-q^2ss'}{4\lambda^{3/2}(s',s,q^2)}
\theta(-q^2s's-m^2\lambda(s',s,q^2)),\qquad q^2<0
\eeq
with $\lambda(s',s,q^2)=(s'+s-q^2)^2-4s's$.
At $q^2=0$ one finds
\beq
\Delta_\pi(s',s,q^2=0)=\frac{s}{8}\sqrt{1-\frac{4m^2}{s}}\;\delta(s'-s)
=\pi\rho_\pi(s)\;\delta(s'-s),
\eeq
and
\beq
\label{ffpiat0}
F_\pi(0)=f_c(0)\int\frac{ds\; G_v^2(s)}{\pi(s -\mpis)^2}\rho_\pi(s)=1
\eeq
As we have mentioned this is just the Ward identity consequence.

One can equivalently formulate the dispersion approach on the light--cone
by introducing the light--cone variables
\beq
k_-=\frac{1}{\sqrt{2}}(k_0-k_z);\quad
k_+=\frac{1}{\sqrt{2}}(k_0+k_z);\quad
k^2=2k_+k_--k^2_\perp;
\eeq
into the integral representation for the \ff spectral density (\ref{deltapi}).
Performing $k_-$ integration
and
setting $(\mu=+)$ in both sides of (\ref{trace1}) one finds

\beq
\label{specden}
\Delta_\pi(s',s,q^2)=\frac1{16\pi}\int \frac{dx d^2k_\perp}{x(1-x)^2}
\delta\left(s-\frac{m^2+k^2_\perp}{x(1-x)}\right)
\delta\left({
s'-\frac{m^2+(k_\perp-xq_\perp)^2}{x(1-x)}
}\right)((s'+s)(1-x)+q^2x)
\eeq
Here we denoted $x=k_{2+}/P_+$ and $k_\perp=k_{2\perp}$.

Substituting (\ref{specden}) into (\ref{ffpi}) and performing $s$ and $s'$
integrations, one derives
\beq
F_\pi(q^2_\perp)=\frac1{\pi}\int dx d^2k_\perp
\psi(x,k_\perp)\psi(x,k_\perp-xq_\perp)\beta(x,k_\perp,q_\perp)
f_c(q^2_\perp),
\eeq
where the radial light-cone wave function of a pion is introduced
\beq
\label{ffpilc}
\psi(x,k_\perp)=\frac{G_v(s)\sqrt{s}}{\pi\sqrt{8}(s-\mpis)\sqrt{x(1-x)}},
\qquad s=\frac{m^2+k^2_\perp}{x(1-x)}
\eeq
$$
\beta=\frac{m^2+k^2_\perp-xk_\perp q_\perp}{\sqrt{m^2+k^2_\perp}\;
\sqrt{ m^2+(k_\perp-xq_\perp)^2}},\qquad \beta(q_\perp=0)=1
$$
The quantity $\beta$ accounts for the contribution of spins.
It is different from unity because both the spin-nonflip and
spin-flip amplitudes of the interacting quark contribute.
The eq.(\ref{ffpiat0}) is the normalization condition
\beq
\int dx\;dk^2_\perp\;|\psi(x,k_\perp)|^2=1.
\eeq

Let us now consider the pion
axial--vector decay constant $f_\pi$.
It is related to the $\pi\to\mu\nu$ decay as
\beq
<\pi^+(P_\pi)|A^{aa}_\mu(0)|0>=i(P_\pi)_\mu\;f_\pi,\quad f_\pi=130\;MeV.
\eeq
To derive the expression for this matrix element we must first consider
the quantity
\beq
<U\bar D|A^{aa}_\mu(0)|0>
\eeq
with $U$ and $D$ constituent quarks
and then single out the pole corresponding to the pion.
Mention that the axial current
$A_\mu(0)=\bar u(0)\gamma_\mu\gamma_5 d(0)$ is defined through current quarks.

\begin{figure}
\begin{center}
\mbox{\epsfig{file=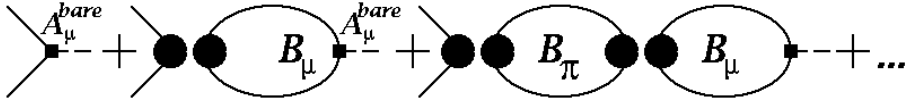,width=12.cm}}
\end{center}
\caption{The series of dispersion graphs for $<Q\bar Q|A_\mu(0)|0>$.
\label{fig:5}}
\end{figure}

The $bare$ matrix element reads ($Q$ denotes a constituent quark)
\beq
<Q(k)\bar Q(P-k)|A^{aa}_\mu(0)|0>_{bare}=\bar Q(P-k)\left[{
\gamma_\mu\gamma_5g^0_A(P^2)+P_\mu\gamma_5h^0_A(P^2)
}\right]Q(k)
\eeq
If current quarks were identical to constituent ones we would have had
$$g_A^0(P^2) \equiv 1, \quad h_A^0(P^2) \equiv 0.$$
It is reasonable to assume that at least $g^0_A(0)$ and $h_A^0(0)$
are not far from these values.
The $bare$ matrix element enters into a single loop graph $B_\mu$
whose spectral density reads
\beq
-\frac{\sqrt{N_c}}{8\pi^2}\int dk_1
dk_2\delta(k^2_1-m^2)\delta(k^2_2-m^2) \delta(P-k_1-k_2)\;
\eeq
$$
\times Sp\left({
[\gamma_\mu\gamma_5 g_A^0(P^2)+P_\mu \gamma_5h_A^0(P^2)]
(\hat k_1+m)\;i\gamma_5(m-\hat k_2)}\right)
$$
So the loop graph is equal to
\beq
B_\mu=iP_\mu[4mg_A^0(P^2)-\frac{P^2}2h^0_A(P^2)]
\sqrt{N_c}\int\frac{ds\; G(s)}{\pi(s-M^2)}\frac{1}{16\pi}\sqrt{1-4m^2/s}
\eeq
After allowing for constituent quark rescatterings we come to the
series of dispersion graphs of Fig.\ref{fig:5} which gives
\beq
<Q(k)\bar Q(P-k)|A^{aa}_\mu(0)|0>=
\bar Q(P-k)\left[{
\gamma_\mu\gamma_5g_A(P^2)+P_\mu\gamma_5h_A(P^2)
}\right]Q(k)
\eeq
with
\beeq
g_A(P^2)= & g_A^0(P^2) & \nonumber \\
h_A(P^2)= &
h^0_A(P^2) &
-\frac{G(P^2)}{1-B_\pi(P^2)}4m[g^0_A(P^2)-\frac{P^2}{2m}h^0_A(P^2)]
\int\frac{ds\; G(s)}{\pi(s-M^2)}\frac{1}{16\pi}\sqrt{1-4m^2/s} \nonumber
\eeeq
The \ff $h_A$ develops a pole at $P^2=m^2_\pi$ as $B_\pi(\mpis)=1$.
Near $P^2=m^2_\pi$ one has
\beq
<\bar QQ|A_\mu|0>=<\bar QQ|\pi>\frac1{m^2_\pi-P^2}<\pi|A_\mu|0>.
\eeq
Comparing the pole terms in (34) and (35) and using the relation
$$
<\pi|Q\bar Q>=\frac{\bar Q i\gamma_5Q}{\sqrt{N_c}}G_v
$$
one finds
\beq
<\pi|A^{aa}_\mu(0)|0>=
iP_\mu 4m[g^0_A(m^2_\pi)-\frac{m^2_\pi}{2m}h^0_A(m^2_\pi)]
\sqrt{N_c}\int\frac{ds\; G_v(s)}{\pi(s-M^2)}\frac{1}{16\pi}\sqrt{1-4m^2/s}
=iP_\mu f_\pi
\eeq
and hence
\beq
\label{weak}
{4m\kappa\sqrt{N_c}}
\int\frac{ds\; G_v(s)}{\pi(s-M^2)}\frac{1}{16\pi}\sqrt{1-4m^2/s}
={f_\pi}
\eeq
with
\beq
\kappa=g_A^0(m^2_\pi)-\frac{m^2_\pi}{2m} h_A^0(m^2_\pi)
\eeq
We can neglect the second term because the small value $h_A^0$ is
further suppressed by $\mpis$.
Finally, one finds
\beq
\label{kappa}
\kappa\approx g_A^0(m^2_\pi)\approx g_A^0(0)\approx 0.75\div 1
\eeq
In terms of the light-cone \wf (\ref{weak}) takes the form
\beq
\kappa \frac{\sqrt{N_c}}{\sqrt2 \pi}
\int{dx\;dk^2_\perp}\frac{m}{\sqrt{m^2+k^2_\perp}}\;\psi(x,k_\perp)={f_\pi}
\eeq

\section{Appendix C: Soft--hard form factor}
\begin{figure}
\begin{center}
\mbox{\epsfig{file=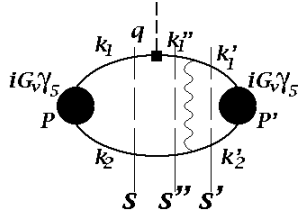,height=3cm}}
\end{center}
\caption{The soft--hard contribution $F^{sh}_\pi$ to pion form factor.
\label{fig:6}}
\end{figure}

The soft--hard \ff $F^{sh}$, given by the graph of Fig.\ref{fig:6},
accounts for the assumption that at $s''>s_0$ the $\bar QQ$ interaction
is described by the convolution of the one--gluon exchange kernel with the
soft--region vertex $G_v$ (the right block in Fig.\ref{fig:6}).
To derive the spectral density of $F^{sh}$, we start with the corresponding
Feynman graph with pointlike vertices
\beq
\label{shF}
2P_\mu(q)\;F^{sh}(q^2)=\frac{Sp(\lambda^A\lambda^A)}{4N_c}g^2
\int \frac{dk}{(2\pi)^4i} \frac{dk'}{(2\pi)^4i}
\eeq
$$
\times \frac1{m^2-k^2}\frac1{m^2-(P-k)^2}\frac1{m^2-(P'-k)^2}
\frac1{m^2-k'^2}\frac1{m^2-(P'-k')^2}
\frac1{\mu^2-(k'-k)^2}
$$
$$
\times Sp\left({ (\hat k_1+\hat q+m)\gamma_\mu(\hat k_1+m)
i\gamma_5(m-\hat k_2)\gamma_\alpha(m-\hat k'_2)
i\gamma_5 (\hat k'_1+m)\gamma^\alpha}\right)
$$
To allow only for two--particle intermediate states we consider the
contribution
of $s$, $s''$, and $s'$ cuts and neglect the contribution of three--particle
intermediate states with cutting the gluon line: three--particle states are
beyond
the scope of our consideration. The resulting spectral density over both
soft $s$ and $s'$ reads
\beq
\label{shden}
\Delta^{sh}(s',s,q^2)=
\theta(s<s_0)\theta(s'<s_0)\;g^2\;C_F
\int\frac{ds''}{s''-\mpis}\theta(s''>s_0)\;
Tr(s,s',s'',q^2)
\eeq
$$
\times\int\frac{dk_1 dk_2 dk''_1}{8\pi^2}\delta(k^2_1-m^2)\delta(k^2_2-m^2)
\delta(k''^2_1-m^2)
\delta(P-k_1-k_2)\delta(P''-k_1''-k_2)\delta(k''_1-k_1-q)
$$
$$
\times\int\frac{dk'_1 dk'_2}{8\pi^2}\delta(k'^2_1-m^2)\delta(k'^2_2-m^2)
\delta(P'-k_1'-k'_2)\frac{1}{m_G^2-(k_2-k'_2)^2}
$$
with $P^2=s,\;P''^2=s'',\;P'^2=s'$.

The quantity
$Tr(s,s',s'',q^2)$ is the dispersion expression for the fermion--loop trace
with all fermions taken on mass shell (for details see the Appendix D)
\beq
Sp\left({ (\hat k_1+\hat q+m)\gamma_\mu(\hat k_1+m)
i\gamma_5(m-\hat k_2)\gamma_\alpha(m-\hat k'_2)
i\gamma_5 (\hat k'_1+m)\gamma^\alpha}\right)=2P_\mu(q)Tr
\eeq
$$
Tr=
2s'\left({\alpha(s'',s,q^2)(s''+s-q^2)+q^2}\right)
-4m^2\left({\alpha(s'',s,q^2)(s'+s-q^2)+q^2)}\right)
$$
where
$$
\alpha(s'',s,q^2)=\frac{-q^2(s''+s-q^2)}{(s''-s)^2-2q^2(s''+s)+q^4}
$$
Use again the light--cone variables (\ref{lcvariables}).
Performing $k_-$ integration and denoting $k=k_2$, $k'=k'_2$,
$x=k_{2+}/P_+$, $x'=k'_{2+}/P_+$, $k_\perp=k_{2\perp}$,
$k'_{\perp}=k'_{2\perp}$ we come to the final expression
\beq
\label{shff}
F^{sh}(q_\perp^2)=4\pi C_F
\int
\frac{dxd^2k_\perp}{16\pi^3x(1-x)^2}
\frac{G_v(s)\theta(s<s_0)}{\frac{m^2+k^2_\perp}{x(1-x)}-\mpis}
\frac{\theta(s''>s_0)}{\frac{m^2+(k_\perp-xq_\perp)^2}{x(1-x)}-\mpis}
\eeq
$$
\times\frac{dx'd^2k'_\perp}{16\pi^3x'(1-x')}
\frac{G_v(s')\theta(s'<s_0)}{\frac{m^2+(k'_\perp-x'q_\perp)^2}{x'(1-x')}-\mpis}
Tr(s,s',s'',q^2)\frac{\alpha_s\({-(k-k')^2}\)}{m_G^2-(k'-k)^2}
$$
where
$$
-(k'-k)^2=\frac{(k'_\perp x-k_\perp x')^2+m^2(x-x')^2}{xx'}
$$
and
$$
s=\frac{m^2+k_\perp^2}{x(1-x)},\quad
s''=\frac{m^2+(k_\perp-xq_\perp)^2}{x(1-x)},\quad
s'=\frac{m^2+(k'_\perp-x'q_\perp)^2}{x'(1-x')}
$$
In (\ref{shff}) the renormalization $g^2\to 4\pi\alpha_s(-(k-k')^2)$
is taken into account.

The distribution amplitude which describes the large-$Q^2$ behavior of the \ff
(see \ref{daphi}) is expressed through $G_v$ as
\beq
\phi(x, s_0)=4m\sqrt{N_c}{\kappa}
\int \frac{ds\;G_v(s)}{16\pi^2(s-\mpis)}\theta\({\frac{m^2}{x(1-x)}<s}\)
\theta(s<s_0)
\eeq

\section{Appendix D: Trace calculation in the soft--hard contribution}
\begin{figure}
\begin{center}
\mbox{   \epsfig{file=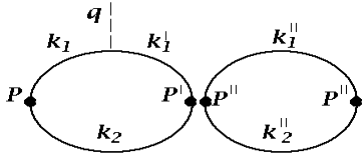,width=5.cm}       }
\end{center}
\caption{Two loops in $F^{sh}$: momentum notation.\label{fig:7}}
\end{figure}

The soft--hard contribution is described by two-loop graph, and the dispersion
technique prescribes that the total momenta squared of the $\bar QQ$ pair
should be taken different for each loop, and then the integration
over these values should be performed. So, we must first make
the Fierz rearrangements to obtain trace calculations related to different
loops. Namely, we group the expressions as follows
$$
2P_\mu(q)Tr\equiv Sp\left({ (\hat k''_1+m)\gamma_\mu(\hat k_1+m)
i\gamma_5(m-\hat k_2)\gamma_\alpha(m-\hat k'_2)
i\gamma_5 (\hat k'_1+m)\gamma^\alpha}\right)
$$
$$
=Sp \left[{ (\hat k''_1+m)\gamma_\mu(\hat k_1+m)
i\gamma_5(m-\hat k_2)}\right]
\gamma_\alpha
\left[
{(m-\hat k'_2)i\gamma_5 (\hat k'_1+m)}\right]
\gamma^\alpha
$$
$$
=\sum\limits_{i=S,V,T,A,P}\quad C_i\times
Sp \left[{ (\hat k''_1+m)\gamma_\mu(\hat k_1+m)
i\gamma_5(m-\hat k_2)O_i}\right]
Sp \left[{(m-\hat k'_2)i\gamma_5 (\hat k'_1+m)O_i}\right]
$$
with $C_S=1, C_V=-\frac12, C_A=\frac12, C_P =-1$.
The second trace is nonzero only for $A$ and $P$ and we find
$$
2P_\mu(q)Tr=\frac12
Sp \left[{ (\hat k_1+\hat q+m)\gamma_\mu(\hat k_1+m)
i\gamma_5(m-\hat k_2)\gamma_5\gamma_\alpha}\right]
Sp \left[{(m-\hat k'_2)i\gamma_5 (\hat k'_1+m)\gamma_\alpha\gamma_5}\right]
$$
$$
-Sp \left[{ (\hat k_1+\hat q+m)\gamma_\mu(\hat k_1+m)
i\gamma_5(m-\hat k_2)\gamma_5}\right]
Sp \left[{(m-\hat k'_2)i\gamma_5 (\hat k'_1+m)\gamma_5}\right]
\equiv 2P_\mu(q)(Tr_A+Tr_P)
$$
Each of the expressions $Tr_i$  ($i=A, P$) is represented as a product of
two factors, related to two different loops (see Fig.\ref{fig:7})
We must use the following relations
$$
k_1+k_2=P, P^2=s,\quad
k''_1=k_1+q, k''_1+k_2=P'', P''^2=s''
$$
for the left loop and
$$
k'_1+k'_2=P', P'^2=s'
$$
for the right one and set $all$ the fermions on mass shell.
This procedure yields
$$
Tr_A=-4m^2[\alpha(s'', s, q^2)(s'+s-q^2)+q^2]
$$
$$
Tr_P=2s'[\alpha(s'', s, q^2)(s''+s-q^2)+q^2]
$$
And the final result reads
$$
Tr=\left[{
2s'\left({\alpha(s'',s,q^2)(s''+s-q^2)+q^2}\right)
-4m^2\left({\alpha(s'',s,q^2)(s'+s-q^2)+q^2)}\right)
}\right]
$$

\end{document}